\documentclass{aa} 
 
\usepackage{natbib}

\usepackage{graphicx}
\usepackage[varg]{txfonts}

\def\Bad  {Ba\,{\small II}}
\def\Srd  {Sr\,{\small II}} 
\def\Feu  {Fe\,{\small I}}
\def\Fed  {Fe\,{\small II}}
\def\Tiu  {Ti\,{\small I}}
\def\Tid  {Ti\,{\small II}}

\def\Nau  {Na\,{\small I}}
\def\Teff  {$T_\mathrm{eff}$}
\def\logg  {$\log g$}

\def\vt    {$\rm v_{t}$}
\def\kms   {$\rm km\,s^{-1}$}

\begin{document}
   \title{
The low Sr/Ba ratio on some extremely metal-poor stars.
\thanks{Based on observations obtained with the ESO Very Large
Telescope at Paranal Observatory, Chile (ID 077.D-0299(A) PI Bonifacio, and ID 078.B-0238(A) PI Spite), and on archive data ID 076.D-0451(A) PI Johnson.  The line list and the abundances line by line for the two stars CS\,29493-090 and HE\,0305-4520 are available at the CDS via anonymous ftp to cdsarc.u-strasbg.fr (130.79.128.5)
or via http://cdsweb.u-strasbg.fr/cgi-bin/qcat?J/A+A/
 }
}

\author {
Spite M. \inst{1}\and
Spite F. \inst{1}\and
Bonifacio P. \inst{1}\and
E. Caffau\inst{1,2}\thanks{MERAC fellow}\and
Fran\c cois P. \inst{1,3}\and
Sbordone L. \inst{4,2,1} 
}

\institute {
GEPI Observatoire de Paris, CNRS, Universit\'e Paris Diderot, F-92195
Meudon Cedex France
\and
Zentrum f\"ur Astronomie der Universit\"at Heidelberg, Landessternwarte, K\"onigstuhl 12, 69117, Heidelberg, Germany
\and
Universit\'e de Picardie Jules Verne, 33 rue St-Leu, 80080 Amiens, France
\and
Millennium Institute of Astrophysics, Universidad Catolica de Chile, Av. Vicu\~na Mackenna 4860, 782-0436 Macul, Santiago, Chile
}

%   \date{Received September 15, 1996; accepted March 16, 1997}

\authorrunning{Spite et al.}
\titlerunning{Sr poor extremely metal-poor stars}

% \abstract{}{}{}{}{} 
% 5 {} token are mandatory
 
  \abstract
 {It has been noted that, in classical extremely metal-poor (EMP) stars, the abundance ratio of two well-observed neutron-capture elements, Sr and Ba, is always higher than  [Sr/Ba] = --0.5, which is the value of the solar r-only  process; however, a handful of EMP stars have recently been found with a very low Sr/Ba ratio.
}   
 { We try to understand the origin of this anomaly by comparing the abundance pattern of the elements in these stars and in the classical EMP stars.
}
{For a rigorous comparison with previous data, four stars with very low Sr/Ba ratios were observed and analyzed in the  same way as in the First Stars Program: analysis within LTE approximation through 1D (hydrostatic) model atmosphere, providing homogeneous abundances of nine neutron-capture elements.
}
   {In  CS 22950-173, the only turnoff star of the sample, the Sr/Ba ratio is,  in fact, found to be higher than the r-only solar ratio, so the star is discarded. The remaining stars (CS 29493-090, CS 30322-023, HE 305-4520) are cool evolved giants. They do not present a clear carbon enrichment, but in evolved giants C is partly burned into N, and owing to their high N abundance, they could still have initially been carbon-rich EMP stars (CEMP). The abundances of Na to Mg present similar anomalies to those in CEMP stars.
The abundance patterns of the neutron-capture elements in the three stars are strikingly similar to a theoretical s-process pattern. 
This pattern could at first be attributed to pollution by a nearby AGB, but none of the stars presents a clear variation in the radial velocity indicating the presence of a companion.
The stellar parameters seem to exclude any internal pollution in a TP-AGB phase for at least two of these stars. 
The possibility that the stars are early-AGB stars polluted during the core He flash does not seem compatible with the theory.
}
{}


\maketitle

%
%---------------------------- Introduction -----------------------
\section{Introduction} \label{intro}

The heavy elements (heavier than Zn) are thought to be built by two main  processes.
The s-process operates by slow neutron capture on seed nuclei on a long time scale and the capture is slow compared to the $\beta$  decay  of the affected nucleus. It often happens  in relatively low-mass stars at the end of their long evolution in their AGB (asymptotic giant branch) phase,  \citep[for example][]{KappelerGB11,BisterzoGS12}.
  
The r-process instead occurs on a very short time scale in violent events not yet clearly identified \citep[e.g.][]{LangankeThi13,AokiSB13}: explosions following the core collapse of massive supernovae, and/or mergings of neutron stars or of black holes, and/or jets,  and/or gamma ray bursts,   etc. These two distinct processes build generally different isotopes of a given heavy element, and different element ratios.

 Since the low-mass stars have a very long lifetime, the matter in the very first phases of the Galaxy could not be enriched by the late production of these low-mass stars in their AGB phase, therefore the heavy elements abundances in this matter must only reflect  the r-process production. 

The stars formed from this primitive matter are considered as key objects for constraining the r-process.
These stars are extremely metal-poor, since at their birth the matter was enriched by a very small number of supernovae. 
Following \citet{BeersChris05}, stars are called extremely metal-poor (EMP),  when [Fe/H] \footnote{In the classical notation, $\rm [X/H] = \log (N_{X} / N_{H})_{star} - \log (N_{X} / N_{H})_{Sun}$ } $\leq -3$.  
Here, for convenience, we extend this definition to stars with $\rm[Fe/H] \leq -2.6$, although most of the stars considered are strictly EMP following the classification of \citet{BeersChris05}. We will consider two classes of EMP stars: the classical EMP stars (which are not carbon-rich) and the carbon-rich EMP stars (CEMP).

$\bullet$ The EMP stars. In the classical EMP stars, the scatter of the enrichment in neutron-capture elements is very large, some are rich in neutron-capture elements like CS\,31082-001 \citep{HillPC02,SiqueiraMelloSB13} 
with a mean enhancement reaching $\rm[r/Fe] \simeq +1.0$ dex, but 
others are poor with $\rm[r/Fe] \simeq -1.0$ dex like HD\,122563 \citep{HondaAI06}. 
Generally, [Eu/Fe] is used to measure the r-process enrichment. In the EMP stars it is not always easy to measure the abundance of Eu, but at these low metallicities there is a very good correlation between  [Eu/Fe] and [Ba/Fe] \citep[see, e.g.,][]{MashonkinaCB10,SpiteSpite14}, and thus [Ba/Fe] can be used to estimate the r-process enrichment.

\begin{figure}[ht] 
 \centering     
\resizebox  {6.5cm}{7.0cm}
{\includegraphics {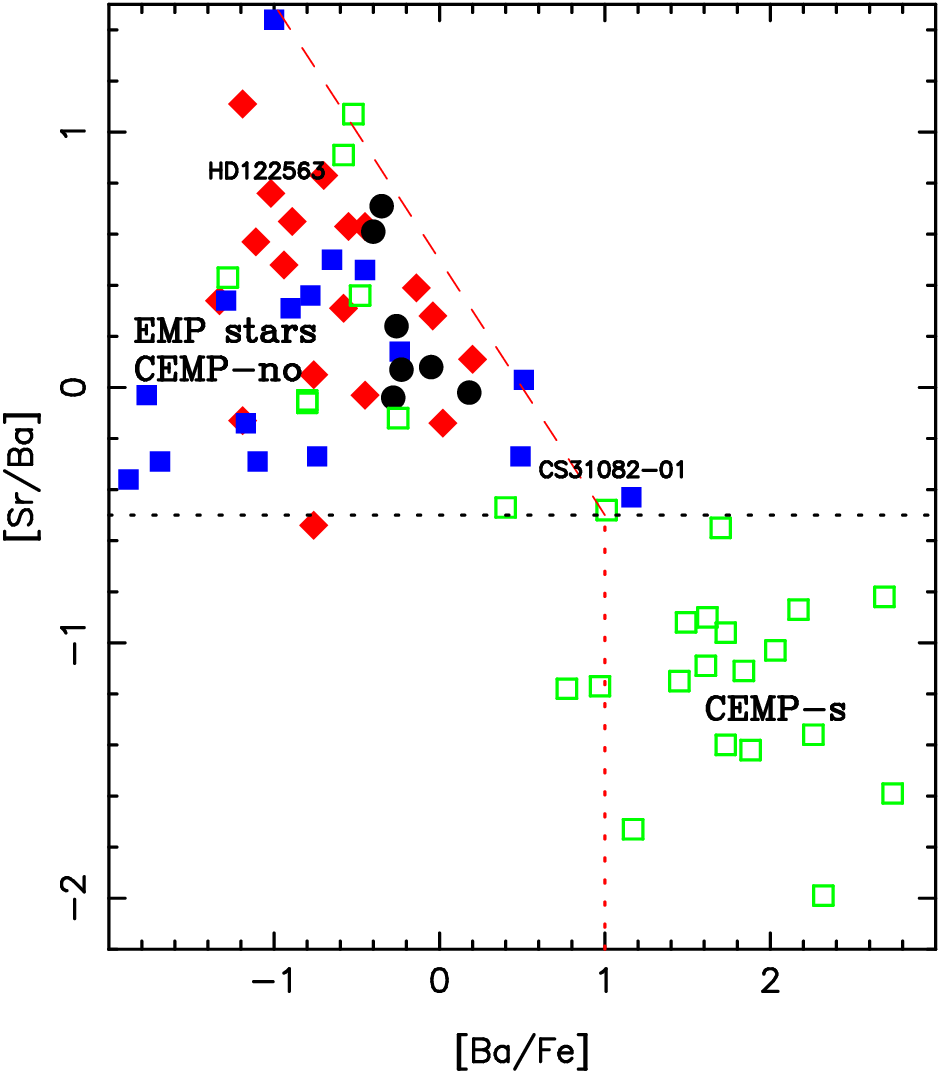} }
\caption[]{[Sr/Ba] vs. [Ba/Fe] for {\bf~~~~~i)} our sample of EMP stars from the Large Program First Stars: filled symbols  (blue squares for unmixed giants, \citet{SpiteCP05}, red diamonds for mixed giants, black circles for dwarfs); and {\bf~~~~~ii}) green open squares for the  CEMP stars in the same range of metallicity gathered from  \citet{AokiRN01,AokiRN02}, \citet{BarbuySS05}, \citet{SivaraniBB06}, \citet{BeharaBL10}, \citet{SpiteCB13}, and \citet{YongNB13}.  The dotted line at [Sr/Ba]=--0.5 represents the value of the r-only solar value of this ratio \citep{MashonkinaG01}.  (Color figures are available in the electronic edition.)\\  
The EMP stars are all located in the left upper corner of the figure, many CEMP stars are located in the same region (they are CEMP-no), but another group (CEMP-s) has a high abundance of Ba: $\rm[Ba/Fe]\gtrsim+1$ associated with a low value of [Sr/Ba] (lower right region).\\
%--The three star symbols represent the stars in Table \ref{abund}. In these stars  $\rm[Ba/Fe]<+1$ as in the classical EMP stars but [Sr/Ba] is very low as in the CEMP-rs stars.  
In the left upper corner, the stars are gathered below the dashed line: [Sr/Ba] =  0.5 -- [Ba/Fe]. } 
\label{scatter1}
\end{figure}

In Fig.\,\ref{scatter1} [Sr/Ba] is plotted vs. [Ba/Fe] for the sample of EMP stars observed in the framework of the ESO Large Program First Stars, hereafter ESO-First stars,  following \citet{HillPC02}, \citet{FrancoisDH07}, and \citet{BonifacioSC09}. All the EMP stars are located in the upper lefthand corner of the figure, and the scatter of the ratio [Sr/Ba] increases strongly when [Ba/Fe] decreases.
All the EMP stars have\\
\indent--$\rm [Ba/Fe] \lesssim +1.0$,\\ 
\indent--$\rm [Sr/Ba] < 0.5 - [Ba/Fe]$,\\
\indent--$\rm [Sr/Ba] \gtrsim -0.5$, and thus greater than the r-only solar value of this ratio: $\rm [Sr/Ba]_{\odot} = -0.5$ following \citet{MashonkinaG01} and \citet{MashonkinaChri14}.
 
It has been shown that, generally speaking, in the EMP stars, the abundance pattern of the neutron-capture elements is not the same when [Sr/Ba] is high (as in CS\,31082-001) and when  [Sr/Ba] is low as in HD\,122563 \citep{HondaAI06,HondaAI07}. It appears also that there are not two distinct populations of EMP stars r-rich and r-poor,  but a continuous evolution of the abundance patterns of the neutron-capture elements with [Ba/Fe] \citep[see also][]{RoedererCK10}.

% Fig. 1

$\bullet$ The CEMP stars.
A significant fraction of the EMP stars is carbon-rich.
\citet{BeersChris05} consider that a star is carbon-rich when 
 $\rm[C/Fe] \geq +1.0$. As a comparison, following \citet{BonifacioSC09}, the mean value of [C/Fe] in {\it normal}  EMP stars (not carbon-rich) is about $\rm +0.40 \pm 0.2$.
The fraction of carbon-enhanced stars increases when the metallicity decreases \citep[see e.g.][]{LucatelloTB05,CarolloFB2014}. They represent 20\% of the stars at $\rm[Fe/H]=-2$, and of the six stars known with $\rm[Fe/H]\leq-4.5$, only one \citep{CaffauBF11,CaffauBF12} does not present a carbon enhancement. 
In Fig.\,\ref{scatter1}, a set of CEMP stars has been gathered from LTE analyses of \citet{AokiRN01,AokiRN02}, \citet{BarbuySS05}, \citet{SivaraniBB06}, \citet{BeharaBL10}, \citet{SpiteCB13} and \citet{YongNB13}.

Several CEMP stars are located in the same region as the EMP stars.  They seem to have the same abundance pattern of neutron-capture elements as the EMP stars \citep{SpiteSpite14}, and are generally not binaries \citep{StarkenburgSC14}.  They are classified CEMP-no \citep[see also][]{CarolloFB2014}.

Other CEMP stars are strongly enriched in neutron-capture elements compared to the EMP stars. In Fig.\ref{scatter1} they have $\rm [Ba/Fe] \gtrsim 1.0$ and $\rm [Sr/Ba] < -0.5$.
It has been shown that these stars have a high value of the ratio [Pb/Eu]: a  "s-process" signature \citep[see, e.g.,][]{MashonkinaRF12,RoedererCK10,SivaraniBM04}. Such a signature may seem unexpected in stars formed so early in the Galaxy. 
But most of these stars have been found to have a variable radial velocity, indicating pollution by a nearby companion.
In fact, the neutron-capture elements observed in these CEMP stars would not reflect the primitive matter that formed the star: they have been produced by the s-process in the higher mass companion of the binary in its AGB phase,  and later transferred to the atmosphere of the (today observed) lower mass companion \citep[see e.g.][]{IzzardGS09}.
In this paper, we will call these stars CEMP-s, without attempting to resolve the CEMP-rs subclass \citep[where Eu is also enhanced, see, e.g.,][]{BarbuySS05} from a CEMP-s subclass where it is not.

Recently, \citet{AokiSB13} have extracted {\it normal} (not carbon-rich) EMP stars ($\rm [Fe/H] \leq -2.5$)  in the SAGA Database \citep{SudaKY08} with measured values of Sr and Ba (260 stars). 
Almost all these stars present a ratio $\rm[Sr/Ba]>-0.5$, as expected from Fig.~\ref{scatter1} for {\it normal} EMP stars.
However they point out that among these stars, a subsample of six EMP stars (not C-rich) had a surprisingly low-[Sr/Ba] ratio, lower than the r-only solar value of this ratio ($\rm[Sr/Ba]=-0.5$).   
In these stars the [Sr/Ba] ratio is similar to the ratio observed in the CEMP-s stars, but they have a ratio $\rm[Ba/Fe]<+1$.  This paper is an attempt to better understand the cause of these anomalies.

%-------------------------- The star sample ------------------------
\section {Observations of the star sample} 

In the subsample of \citet{AokiSB13}, two stars are suspected of being binaries. A pollution by a companion could thus be suspected of being responsible for their abundance anomalies. But four EMP stars are found to have a ratio $\rm[Sr/Ba]\ll -0.5$ and $\rm[Ba/Fe]<+1$ and are, to date, not suspected of being binaries.
%while not  suspected to be binaries. 
These stars are presented in Table \ref{starsAoki}.

%TAB 1
\begin{table}[ht]
\begin{center}    
\caption[]{Stars selected by \citet{AokiSB13} with $\rm [Fe/H] \leq -2.5$,  $\rm [Ba/Fe]< 1.0$ and $\rm [Sr/Ba] \ll -0.5$ }
\label{starsAoki}
\begin{tabular}{l@{~~~}c@{~~~}c@{~~~}c@{~~~}c@{~~}r@{~~}c@{}c@{}c }
 Star        & [Fe/H]& [Ba/Fe] & [Sr/Ba]& \Teff & \logg & Ref\\
\hline
CS~22950-173 & --2.50	&--0.04   & --0.72  & 6800  &  4.5  & 1\\
CS~29493-090 & --2.82	& +0.52   & --1.41  & 4700  &  1.3  & 2\\
CS~30322-023 & --3.40	& +0.54   & --1.05  & 4100  &--0.3  & 3\\
~~~~~~''~~~  & --3.26	& +0.59   & --1.10  & 4300  &  1.0  & 4\\
HE~0305-4520 & --2.91	& +0.59   & --1.25  & 4820  &  1.3  & 2\\
\hline
\multicolumn{7}{l}{1- \citet{PrestonSne00}}\\
\multicolumn{7}{l}{2- \citet{BarklemCB05}}\\
\multicolumn{7}{l}{3- \citet{MasseronEF06}}\\
\multicolumn{7}{l}{4- \citet{AokiBC07}}\\
\end{tabular}
\end{center}
\end{table}

%TAB 2
\begin{table}
\begin{center}    
\caption[]{Radial velocities (geocentric and barycentric) of the three stars studied. The precision of the radial velocities is about 1\,\kms.} 
\label{VVr}
\begin{tabular}{l@{~~~}c@{~~~}r@{~~~}r@{~~~}r@{~~}c@{~~}c@{}c@{}c }
 Star        & MJD       &  RV     & RV     &   RV  \\ 
             &           & (geo)   & corr   & (bary)\\
\hline
CS~22950-173 & 53848.34  &  +38.7  &  +29.8 &  +68.5\\
\\
CS~29493-090 & 53637.17  & +287.9  & --18.8 & +269.1\\
CS~29493-090 & 53645.18  & +291.5  & --21.7 & +269.8\\
CS~29493-090 & 53645.22  & +291.85 & --21.75& +270.1\\
\\
HE~0305-4520 & 54024.25  & +137.2  & --0.6  & +136.6\\
\hline
\end{tabular}
\end{center}
\end{table}

The star CS\,30322-023 has been studied in detail by \citet{MasseronEF06,MasseronJP10} and \citet{AokiBC07}.
We  analyzed high-resolution spectra for the three other stars. 
The spectra of CS\,22950-173 and HE\,0305-4520 were obtained at the VLT telescope with the UVES spectrograph \citep{DekkerDK00} in the course of our own observing programs. The UVES spectra of CS\,29493-090 (HE\,2156-3130) were retrieved from the ESO archive.\\
The resolving power of the spectra is about 40,000 with five pixels per resolution element. The spectra cover the ranges $\rm 330<\lambda <451\,nm $ (blue arm) and  $\rm 480<\lambda <680\,nm $ (red arm).
The spectra were reduced (optimum extraction, division by the flat field, wavelength calibration) using the standard UVES pipeline.   
The S/N of the spectra (per pixel) at 420nm is 110 for CS~22950-173, 80 for CS~29493-090 and 100 for HE~0305-4520.

Since the existence or non-existence of a companion is essential for explaining the peculiar heavy element patterns in metal-poor stars, we carefully measured the radial velocity of the three new low-Sr/Ba stars (Table \ref{VVr}). The measurements were done on the blue spectra, and the precision of the measurements is about 1~ \kms. These measurements are discussed in sections \ref{to} and \ref{rvgi} and compared to the values found in the literature.

\section {Analysis}

We carried out a classical LTE analysis, which is homogeneous with the analysis of the ESO-First stars sample: red giant branch (RGB) stars  \citep{CayrelDS04} and turnoff stars \citep{BonifacioSC09},  using OSMARCS model atmospheres \citep{GustafssonBE75,GustafssonEE03} and the {\tt turbospectrum} spectral synthesis code \citep{AlvarezP98,Plez12}. 

For the two giants,  the effective temperatures \Teff ~were taken from \citet{BarklemCB05} and deduced from colors. For the turnoff star CS~22950-173, we adopted a temperature based on 3D profiles of the $\rm H\alpha$ wings taking into account  the influence of the gravity  on this profile iteratively (Sbordone et al., in preparation). This temperature is a little higher than the value adopted by \citet{SbordoneBC10} but a little lower than the temperature adopted by \citet{PrestonSne00}.\\

The microturbulence velocity was derived from the \Feu ~lines, requiring that the abundance derived for individual lines be independent of the equivalent width of the line, and the surface gravity \logg ~was determined by requiring that the Fe and Ti abundances derived from neutral and ionized lines be the same.

The adopted parameters are given in Table \ref{models}. These parameters are in the same range as the models adopted for the ESO-First stars sample, giant and turnoff stars \citep{CayrelDS04,BonifacioSC09} and lead to the same uncertainties: about 100K on \Teff, 0.2 dex on \logg, and 0.2\,\kms~ on \vt,  which induces an uncertainty of about 0.1 dex on [Fe/H].

%TAB 3
\begin{table}[ht]
\begin{center}    
\caption[]{Parameters of the models adopted for the three stars analyzed here and the corresponding errors} 
\label{models}
\begin{tabular}{l@{~~~}c@{~~~}r@{~~~}c@{~~~}c@{~~}c@{~~}c@{}c@{}c }
 Star        & \Teff & \logg & \vt & [Fe/H]\\
\hline
CS~22950-173 & 6615  &  4.1  & 1.4  &--2.6	\\
CS~29493-090 & 4700  &  1.3  & 1.9  &--3.1	\\
HE~0305-4520 & 4820  &  1.3  & 2.0  &--3.0	\\
\hline
\end{tabular}
\end{center}
\end{table}

% Fig. 2
\begin{figure}[ht]
 \centering 
\resizebox  {8.5cm}{3.8cm}
{\includegraphics {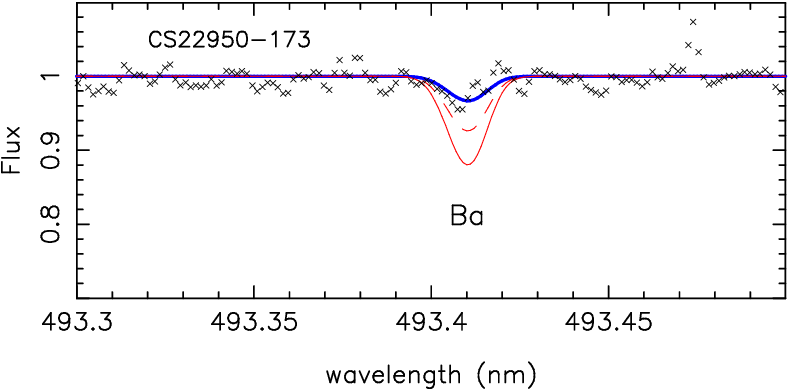} }
\caption[]{Profile of the 493.4 \Bad ~lines in CS\,22950-173 computed with our adopted model (Tab.\ref{starsAoki}) and [Ba/Fe]=--0.53 (adopted value, thick blue line), and [Ba/Fe]=--0.04 (thin red line). The dashed line corresponds to a profile of the Ba line computed with the model adopted by \citet{PrestonSne00} and [Ba/Fe]=-0.04.} 
\label{ba22950}
\end{figure}

\section{The turnoff star CS~22950-173}  \label{to}
In our set of stars, CS~22950-173 is the only turnoff star, and the chemical composition of the atmosphere of this star could not be affected by a mixing with deep layers.
Following \citet{AokiSB13}, this star has been found to be a binary by \citet{PrestonSne00}, suggesting that 
an anomaly in its abundance pattern could be due to a mass transfer from a now defunct companion, when it was in its AGB phase. However, we checked that \citet{PrestonSne00} do not indicate in their Table 1 that this star is suspected to be a binary, and it is even listed  (their Table 4) among the ``blue metal-poor stars with constant radial velocities''.
Moreover, in our spectrum obtained in 2006, we measured a barycentric radial velocity (Table \ref{VVr}), in good agreement with the value ($=69.4 \pm 1$ \kms) measured by \citet{PrestonSne00} on spectra obtained between 1993 and 1998.

A first analysis of this star has been published in \citet{SbordoneBC10} giving the abundances of Fe and Li. A new more complete analysis will be published soon by Sbordone et al. (in preparation). The CH band is very weak, and we found  $\rm[C/Fe] \leq +0.78$: the star is not a CEMP star, according to the definition of \citet{BeersChris05}.

We could measure two \Srd ~lines at 407.771 and 421.552nm and two \Bad ~lines at 493.408 and 614.171nm, and we obtained [Sr/H]=--3.24  and [Ba/H]=--3.13 ([Ba/Fe] = --0.53) i.e. [Sr/Ba]=--0.11.\\ 
In Fig.\ref{ba22950} we compare the observed spectrum in the region of the Ba   line at 493.408nm with\\
(i) synthetic profiles computed with our adopted model and [Ba/Fe]=--0.53 (adopted value) and [Ba/Fe]=--0.04 (Preston \& Sneden value), and\\ 
(ii) the synthetic profile computed with  the model of \citet{PrestonSne00} and [Ba/Fe]=--0.04.

It can be deduced from this figure that the value of [Ba/Fe] measured by Preston \& Sneden is not compatible with our spectrum and the model we adopted. If the model of \citet{PrestonSne00} had been adopted, a higher value of [Ba/Fe] would be found ($\rm[Ba/Fe] \simeq -0.35$) but also a higher value of [Sr/Fe] and thus finally, the ratio [Sr/Ba] would remain unchanged.
Finally, the star CS~22950-173, with $\rm[Sr/Ba]  \simeq -0.1$, is not a low-Sr/Ba star, and we will no more consider this star in the following. 
%(it is compatible with a production of heavy elements 
%dominated by the early r-process completed with some production of C and s-process elements.

%Table 4
\begin{table}
\begin{center}    
\caption[]{LTE abundances in low-[Sr/Ba] ratio giant stars. For Na we give also the NLTE value. 
The data for CS\,30322-023 are taken from \citet{MasseronEF06}}. 
\label{abund}
\begin{tabular}{l@{~~~}r@{~~~}r@{~~~}r@{~~~}r@{~~}r@{~~}r@{}c@{}c }
 & \multicolumn{2}{l}{\small{CS\,30322-023}} & \multicolumn{2}{l}{\small{CS\,29493-090}}  & \multicolumn{2}{l}{\small{HE\,0305-4520}} \\
\hline 
$\rm [Fe/H]$           &\multicolumn{2}{c}{--3.40}&\multicolumn{2}{c}{--3.10}&\multicolumn{2}{c}{--2.95}\\
$\rm^{12}C/^{13}C$     &\multicolumn{2}{c}{4}&\multicolumn{2}{c}{5.5}&\multicolumn{2}{c}{4}\\
\hline
                 &[X/Fe]&$\sigma$&~~~[X/Fe]&$\sigma$  &~~~[X/Fe]&$\sigma$\\
$\rm C_{(CH)}$   &    0.80  &  - &    0.73  & 0.10    &  0.42  & 0.10    \\
N                &    2.91  &  - &    1.51  & 0.15    &  1.58  & 0.15    \\
C+N              &    1.70  &  - &    1.03  &	 -    &  0.99  &  -      \\
O                &    0.63  &  - &     -    &	 -    &  0.83  & 0.10    \\ 
Na               &    1.29  &0.24&    1.13  & 0.10    &  1.04  & 0.10    \\
$\rm Na_{(NLTE)}$&    0.79  &0.24&    0.53  & 0.10    &  0.54  & 0.10    \\
Mg               &    0.80  &0.10&    0.98  & 0.13    &  0.36  & 0.11    \\
Ca               &    0.30  &0.12&    0.27  & 0.10    &  0.26  & 0.10    \\
\Tiu             &  --0.20  &0.19&    0.11  & 0.05    &  0.16  & 0.05    \\
\Tid             &    0.23  &0.13&    0.17  & 0.05    &  0.21  & 0.05    \\
\Feu             &    0.01  &0.18&  --0.04  & 0.05    &--0.01  & 0.05    \\
\Fed             &  --0.01  &0.19&    0.03  & 0.05    &  0.04  & 0.05    \\
Sr               &  --0.50  & -  &  --1.02  & 0.11    &--0.76  & 0.11    \\   
Y                &  --0.36  &0.15&  --0.82  & 0.10    &--0.58  & 0.10    \\   
Zr               &    0.14  &0.30&  --0.50  & 0.10    &--0.38  & 0.11    \\   
Ba               &    0.52  &0.10&    0.43  & 0.10    &  0.32  & 0.10    \\   
La               &    0.46  &0.10&    0.23  & 0.12    &  0.09  & 0.12    \\   
Ce               &    0.59  &0.24&    0.19  & 0.10    &  0.31  & 0.12    \\   
Nd               &    0.57  &0.29&    0.23  & 0.11    &  0.32  & 0.10    \\   
Eu               &  --0.63  & -  &  --0.21  & 0.11    &--0.31  & 0.11    \\   
Dy               &  --0.13  & -  &     -    &	 -    &     -  &  -      \\ 
Pb               &    1.49  &0.20&    1.15  & 0.18    &	1.25  & 0.22     \\   
\hline
\end{tabular}
\end{center}
\end{table}

\section{The giant stars}
The abundances of the elements in the three low-Sr/Ba giants, CS\,29493-090, HE\,0305-4520 (this paper), and  CS\,30322-023 \citep{MasseronEF06,MasseronJP10} are given in Table \ref{abund}  with the stochastic errors arising from random uncertainties in the oscillator strengths and in the measurement of the equivalent widths or the profiles\footnote{The line list, and the abundances line by line are available in electronic form at the CDS via anonymous ftp to cdsarc.u-strasbg.fr (130.79.128.5) 
or via http://cdsweb.u-strasbg.fr/cgi-bin/qcat?J/A+A/}. Systematic uncertainties are mainly due to the adopted stellar parameters. The total uncertainty can be estimated as the quadratic sum of the stochastic and the systematic errors. Because of the similarity of the response of a given set of elements to changes in stellar parameters, systematic errors largely cancel out, reducing the uncertainty on the relative abundances. In the range of temperature gravity and metallicity considered here, the systematic error on [X/Fe] is less than 0.1  \citep[see][]{HillPC02,CayrelDS04,FrancoisDH07}.

 It is interesting to note that in CS\,29493-090 (Table \ref{abund}) the abundance of some elements (C, N, Na, Mg, Ba, Fe) has been determined independently by \citet{MasseronJL12}. They used a model that is only slightly different (\Teff=4700K, \logg=0.9, \vt=1.6 \kms), but the agreement in the determination of [Fe/H] and [X/Fe] is good with a $\rm \Delta [X/Fe] < 0.15$ dex. The main difference is observed for [N/Fe] because Masseron et al. only used  the CN band, while we used the CN and the NH bands. If we had only used the CN band, we would have found [N/Fe]=+1.34, and the difference $\rm \Delta [N/Fe]$ would drop down to 0.11 dex.

Below we compare the abundances of the elements in these three stars,

(i) to our set of classical EMP stars studied in the same way in the frame of the ESO-First stars program \citep{HillPC02,CayrelDS04,SpiteCP05,BonifacioSC09},

(ii) to CEMP stars, observed and analysed also homogeneously in the frame of the ESO-First stars program: CS\,22892-052 \citep{SnedenCL03,CayrelDS04}, CS\,22949-037 \citep{McWilliamPS95,DepagneHS02}, and CS\,31080-095, CS\,22958-045, CS\,29528-041 \citep{SivaraniBB06}, and to CEMP stars in the same range of metallicity, gathered from  the literature \citet{AokiRN01,AokiRN02}, \citet{BarbuySS05}, \citet{BeharaBL10}, \citet{SpiteCB13}, \citet{YongNB13}.
We have divided the CEMP stars in two subclasses:  the CEMP-s, with $\rm[Ba/Fe]>1.0$  and the CEMP-no  with $\rm[Ba/Fe]\leq 1.0$, (see Fig. \ref{scatter1}).
%
%CS\,22892-052 and CS\,22949-037 do not present anomaly of the heavy elements abundances. CS\,22892-052 has the same abundance pattern of heavy elements as the classical EMP star CS31082-001 while CS\,22949-037 seems to have a pattern similar to HD\,122563 \citep[see][]{SpiteSpite14}.

% Fig. 3
\begin{figure}[ht]
 \centering  
\resizebox  {8.8cm}{4.8cm}
{\includegraphics {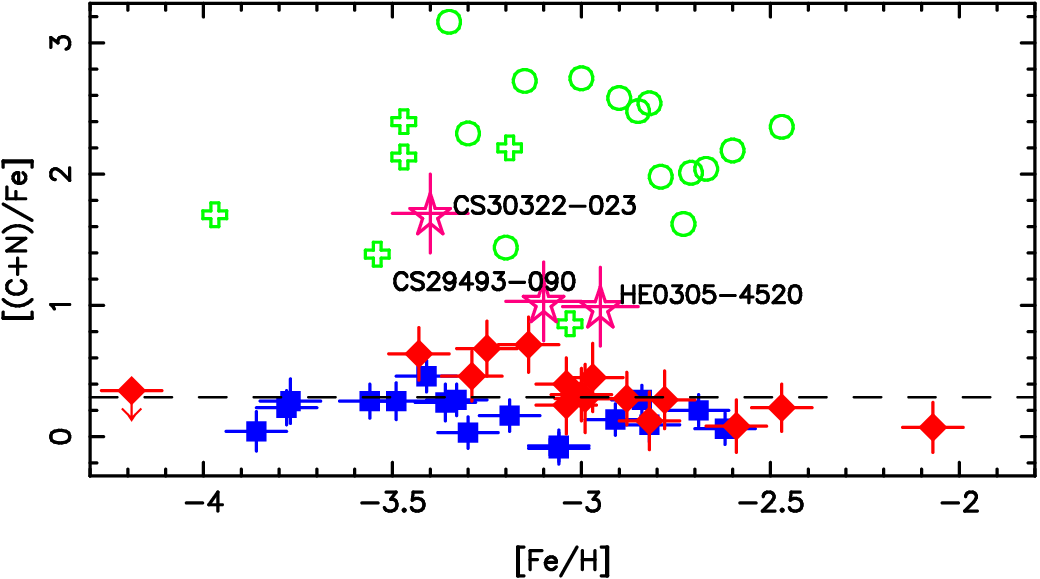} }
\caption[]{Comparison of the abundance of [(C+N)/Fe]. ~~~(i) For the sample of EMP stars studied in the framework of the large program "First Stars": filled blue squares for unmixed RGB stars, filled red diamonds for mixed RGB's, and filled black circles for the turnoff stars (no EMP turnoff stars appear in this figure, because it was not possible to measure the nitrogen abundance in these stars \citep{BonifacioSC09});
~~~(ii) for CEMP-no (open green crosses);
~~~(iii) for CEMP-s (open green circles), and ~~~(iv) for the three low-[Sr/Ba] giants here studied (open pink star symbols).\\ 
These three low-[Sr/Ba] stars are located in the region generally occupied by the carbon-rich stars.
}    
\label{cpn}
\end{figure}

% 5.1
\subsection{C N O abundances} \label{parcno}

With a ratio [C/Fe] between 0.42 and 0.80 dex (Table \ref{abund}), the three low Sr/Ba stars are not CEMP stars according to the definition of \citet{BeersChris05}: $\rm [C/Fe] > +1$. 

But in the atmosphere of giant stars, AGB, or even red giant branch (RGB) stars, the carbon abundance does not always represent the abundance of C in the gas that formed the star.
It has been shown, in particular, that in classical EMP stars evolving along the RGB, an extra mixing occurs when a star crosses the bump in the luminosity function \citep[see:][]{SpiteCP05,SpiteCH06}.
At $\rm[Fe/H] \approx -3.0$, the extra mixing occurs for a temperature of about 4800K and a (spectroscopic) gravity $ \approx 1.5$. 
The atmosphere is mixed with CNO processed material: [C/Fe] decreases and [N/Fe] increases,  but [(C+N)/Fe] remains (almost) constant with a mean value close to +0.3dex (see Fig. \ref{cpn}).

The three low-Sr/Ba stars are cool giants (Table \ref{models}),  and thus it is possible that in their atmosphere, part of the original carbon has been transformed into nitrogen. As a consequence, it is more suitable to compare the abundance of C+N in the low-Sr/Ba stars to the abundance of C+N in classical EMP stars and  in CEMP  stars  (Fig.\ref{cpn}). 
 
As for the CEMP-s or the CEMP-no stars, the three low-Sr/Ba stars, present a ratio [(C+N)/Fe] higher than the classical EMP stars (Fig.\ref{cpn}). It is interesting to note that the CEMP-s and the CEMP-no stars occupy different regions of the figure, the CEMP-s having, at a given metallicity, a higher value of [(C+N)/Fe] than the CEMP-no. This difference could reflect the different origins of the overabundance of C and N in CEMP-s and CEMP-no stars \citep[see also,][]{SpiteCB13}.

The three low-Sr/Ba stars are located in the region of the CEMP-no stars. They could have been born carbon-rich and later, could have undergone an extra mixing, which would explain their relatively low carbon abundance associated with a high value of the abundance of nitrogen.
This extra mixing is also characterized by a low value of the $\rm ^{12}C/^{13}C$ ratio (close to the equilibrium value), which is indeed observed in these three stars  (see upper part of Table \ref{abund}).

% Fig. 4 O
\begin{figure}[ht]
 \centering
\resizebox  {8.8cm}{4.8cm}
{\includegraphics {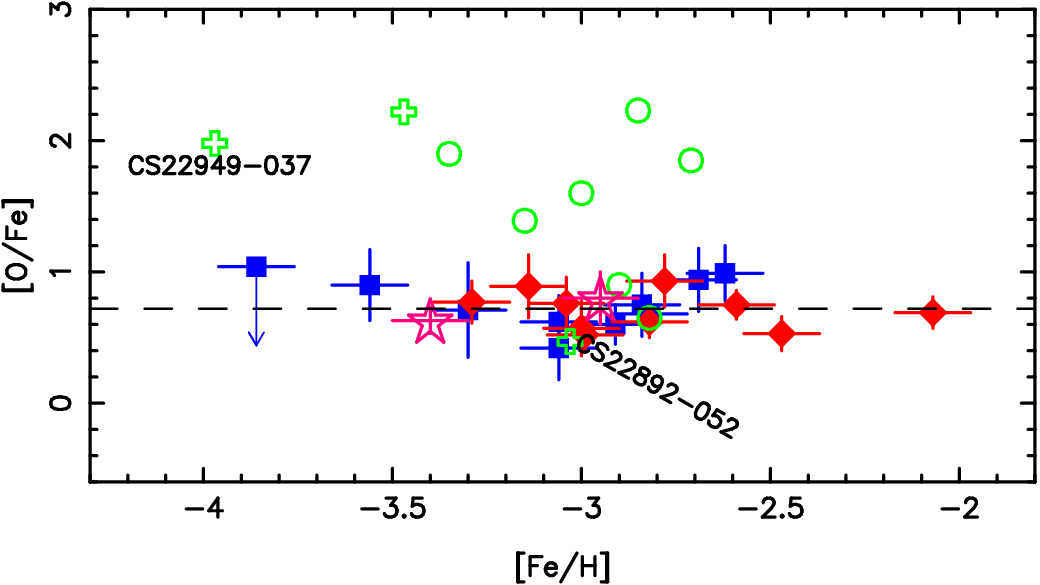} }
\caption[]{Comparison of the abundance of oxygen in a sample of RGB stars studied in the framework of the ESO large program "First Stars", in CEMP-s and CEMP-no stars and in the low-Sr/Ba giants. The symbols are the same as in Fig. \ref{cpn}. The three low-Sr/Ba  giants have an oxygen abundance similar to the abundance measured in the classical EMP stars, but several CEMP stars also have a {\it normal} oxygen abundance.} 
\label{oxy}
\end{figure}

In two of these stars the oxygen abundance could be derived from the forbidden line at 630~nm, and when compared to the set of normal EMP stars this abundance appears to be normal, although in some carbon-rich stars (but not all), it is very high  (see Fig. \ref{oxy}).

In Figures \ref{cpn} and \ref{oxy}, only the giants of the large program First Stars appear, since it has not been possible to measure N and O in the subsample of turnoff stars.

% Fig. 5 Na
\begin{figure}[ht]
 \centering
\resizebox  {8.8cm}{4.8cm}
{\includegraphics {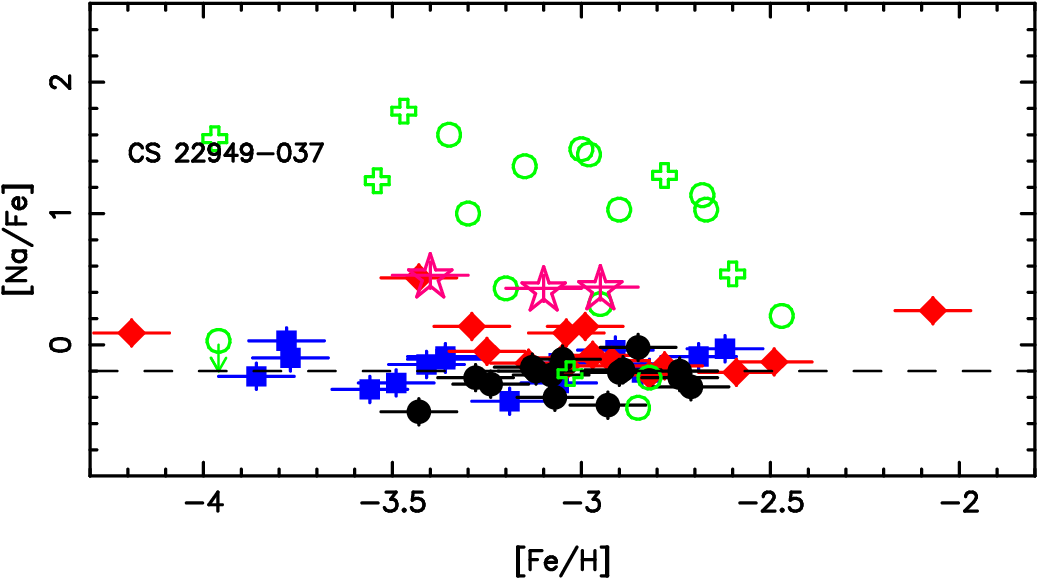} }
\caption[]{Comparison of the abundance of sodium in a sample of RGB stars studied in the framework of the ESO large program "First Stars" and in the three low-[Sr/Ba] giants studied here. The symbols are the same as in Fig. \ref{cpn}.  The three low-Sr/Ba giants have a high sodium abundance like several RGB mixed stars, but [Na/Fe] in these stars is less than the value observed in the majority of the CEMP-s or CEMP-no stars.} 
\label{na}
\end{figure}

% Fig. 6 Mg
\begin{figure}[ht]
 \centering     
\resizebox  {8.8cm}{4.8cm}
{\includegraphics {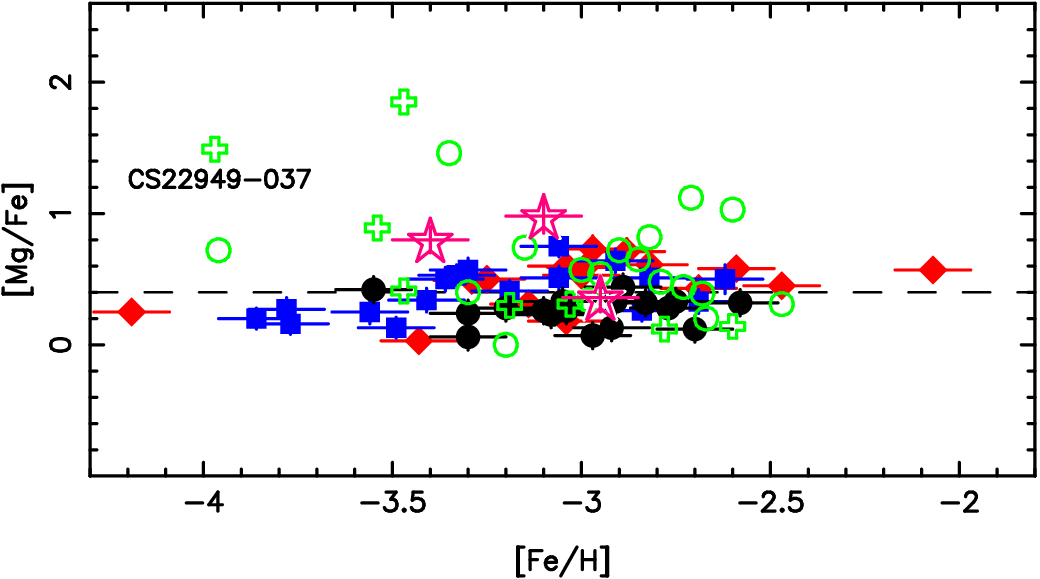} }
\caption[]{Comparison of the abundance of magnesium in the ESO-First stars  and in the three low-Sr/Ba giants studied here. The symbols are the same as in Fig. \ref{cpn}. In two low-Sr/Ba giants, [Mg/Fe] is  higher than in the classical EMP stars as in many CEMP-s or CEMP-no stars,  but in the third one [Mg/Fe] is close to the mean value, as it is also in many CEMP-s or CEMP-no stars.} 
\label{mg}
\end{figure}

% 4.2
\subsection{The light elements from Na to Mg}  \label{parnaca}

Since in these very metal-poor stars,  the sodium abundance is deduced from the  \Nau~ resonance lines, the NLTE correction  can reach --0.5~dex, and it is different in turnoff and giant stars.
As a consequence, in Table \ref{abund} and in Fig. \ref{na}, in order to allow a comparison with the Na abundance in the classical EMP and CEMP stars,  the $\rm[Na/Fe]_{NLTE}$ has been estimated from the computations of \citet{GehrenLS04} and \citet{AndrievskySK07}.
 
In Fig. \ref{na} and \ref{mg} we compare the sodium and magnesium abundances in classical EMP, CEMP-s, CEMP-no stars and in the three peculiar low-Sr/Ba stars. The Na abundance in these three stars is high (as in several mixed RGB stars), but not as high as in most CEMP stars (CEMP-s or CEMP-no).
The slight overabundance of magnesium in these stars is {\it normal} for EMP or CEMP stars, but [Mg/Fe] is rather high in two of these stars  (Fig. \ref{mg}), but not as high as in many CEMP-s or CEMP-no stars.\\

The classical EMP stars present a very low scatter of [C+N/Fe], [O/Fe], [Na/Fe], [Mg/Fe]. This low scatter is illustrated in Fig. \ref{cpn}, \ref{oxy}, \ref{na}, \ref{mg} (filled symbols). The pattern of the abundances of the elements from C to Mg in CS\,30322-023, CS\,29493-090, and HE\,0305-4520 shows that these stars are not classical EMP stars and it allows us to classify them rather among the CEMP stars. The abundances of these light elements in the CEMP-s or the CEMP-no stars is so scattered that it is not possible at this stage to constrain the origin of the overabundance of C+N, Na, and Mg better in the three low-Sr/Ba stars.

% Fig. 7  ecart Sr/Ba
\begin{figure}[ht]
 \centering     
\resizebox  {6.1cm}{6.5cm}
{\includegraphics {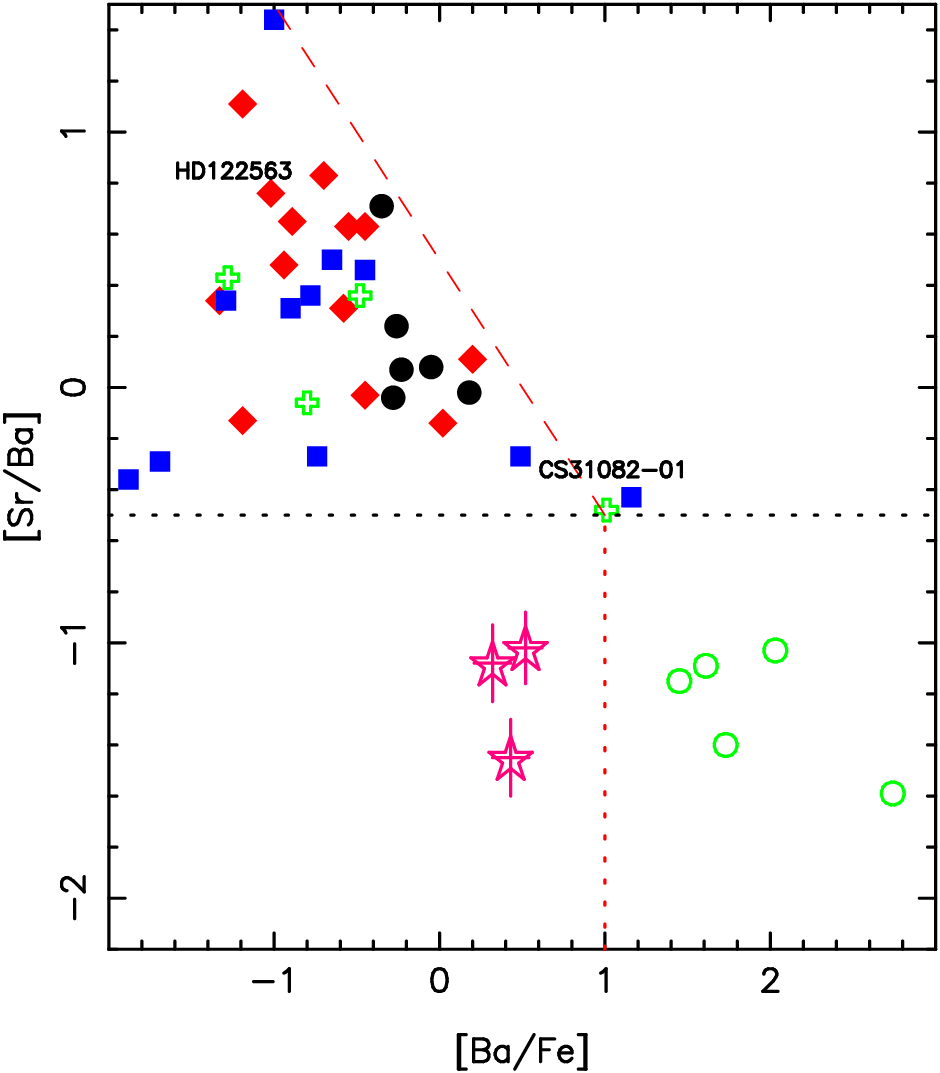} }
\caption[]{[Sr/Ba] vs. [Ba/Fe] for {\bf~~~~i)} our sample of EMP stars (Large Program First Stars restricted to the stars with $\rm -3.5 \leq [Fe/H] \leq -2.7$), for {\bf~~~~~ii)} the  CEMP stars in the same range of metallicity gathered from  \citet{AokiRN01,AokiRN02}, \citet{DepagneHS02}, \citet{BarbuySS05}, and \citet{YongNB13}, and for {\bf~~~~~iii)} the three low-Sr/Ba giants (this paper).
Symbols are the same as in Fig. \ref{cpn}.\\
The three low-Sr/Ba giants occupy a very peculiar position in this diagram: they associate a low value of [Sr/Ba] like in the CEMP-s stars to a relatively low value of [Ba/Fe] ($\rm[Ba/Fe]<1$) like in the classical EMP stars or the CEMP-no.\\
%--The three star symbols represent the stars in Table \ref{abund}. In these stars  $\rm[Ba/Fe]<+1$ as in the classical EMP stars but [Sr/Ba] is very low as in the CEMP-rs stars.  
} 
\label{scatter2}
\end{figure}

% Fig. 8 comp r, s
\begin{figure}[ht]
 \centering     
\resizebox  {8.0cm}{3.36cm}
{\includegraphics {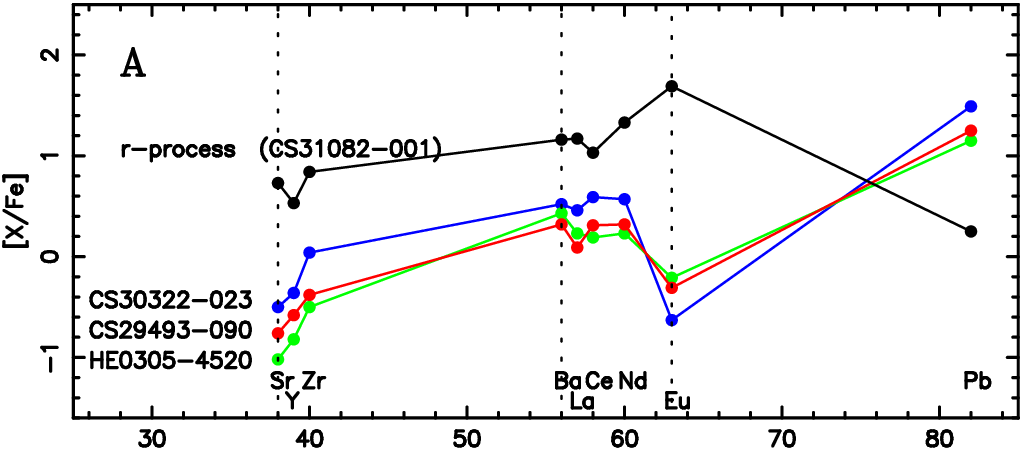} }
\resizebox  {8.0cm}{3.36cm}
{\includegraphics {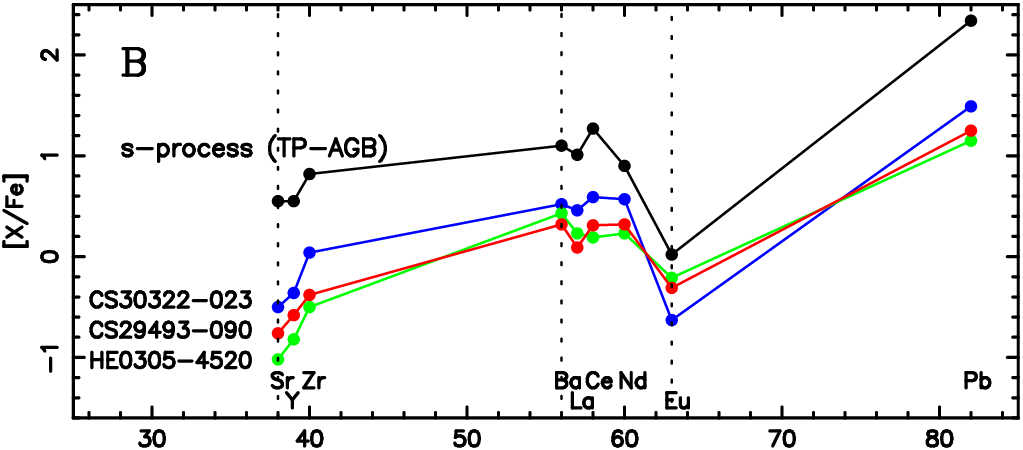} }
%\resizebox  {8.0cm}{3.72cm}
%{\includegraphics {heavy-sur-Fe-coreHe.ps} }
\caption[]{Neutron-capture element ratios relative to iron [X/Fe], in the three low-Sr/Ba stars (CS\,30322-023, blue line; CS\,29493-090, red line; HE\,0305-4520, green line), are compared ~~~~~in A) to the ratios observed in CS\,31082-001, a typical example of the  r-process, ~~~~~in B) to the s-process computations in an AGB star during the thermal pulses shifted by --1.4 dex. 
%and ~~~~~in C) to the s process computations during the core He flash shifted by --4 dex.
} 
\label{surfe}
\end{figure}

% Fig. 9 comp pattern
\begin{figure}[h]
 \centering     
\resizebox  {6.8cm}{4.1cm}
{\includegraphics {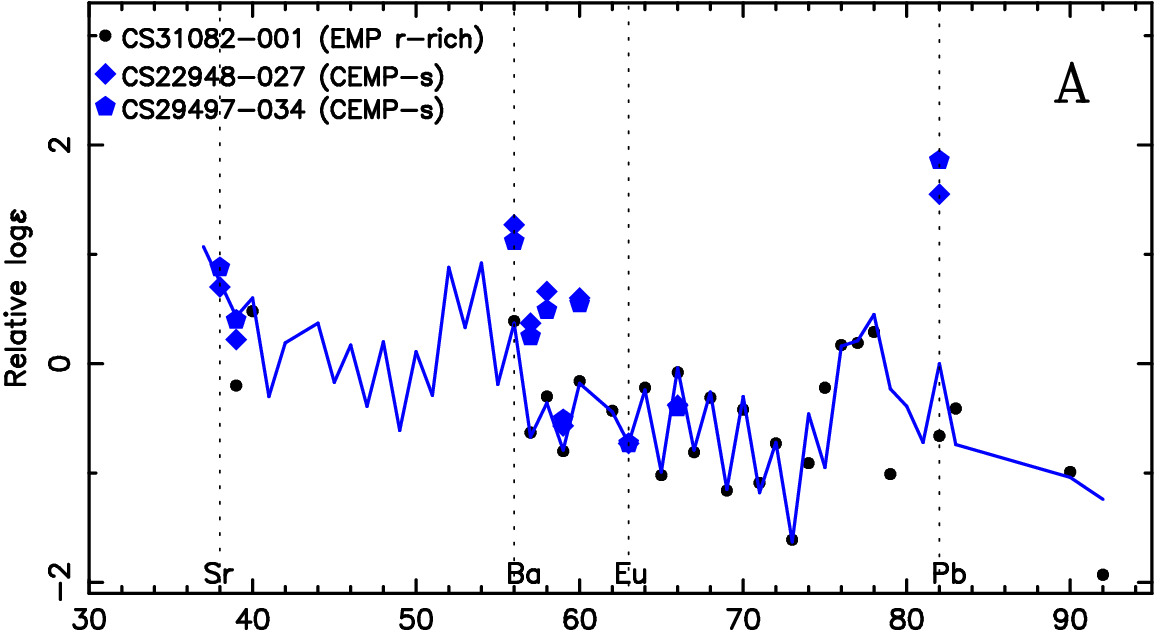} }
\resizebox  {6.8cm}{4.1cm}
{\includegraphics {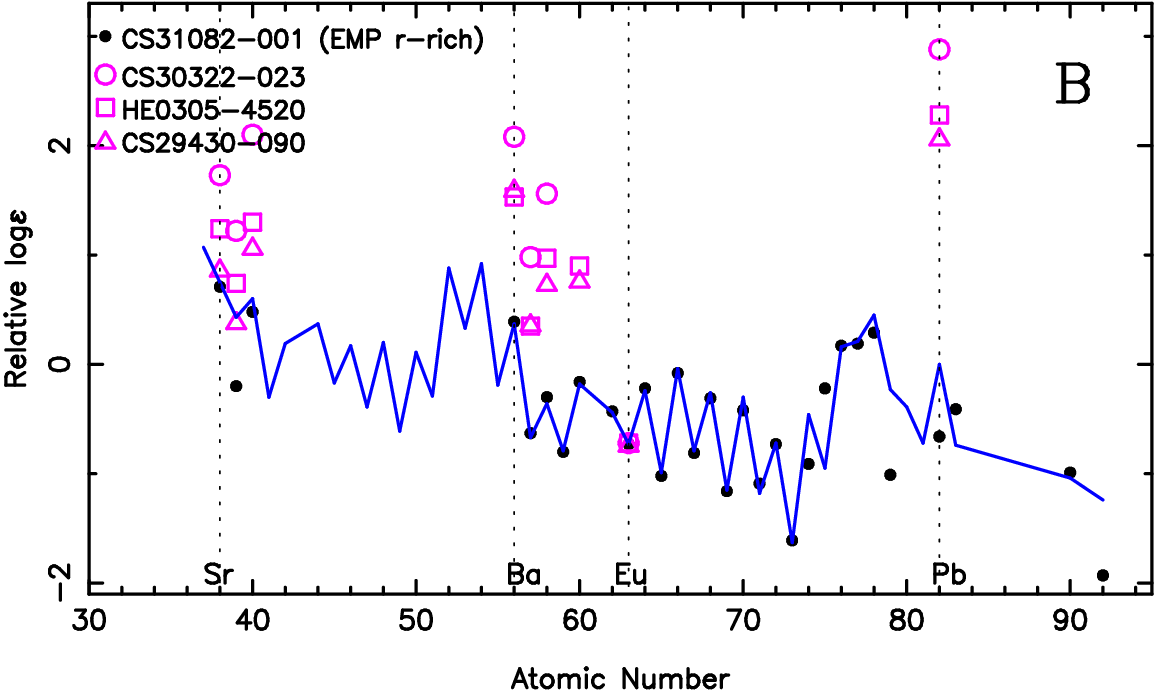} }
\caption[]{
Panel A: abundance pattern of the heavy elements in two classical CEMP-s stars CS\,22948-027 and CS\,29497-034 (blue diamonds and blue pentagons) and compared to the abundance pattern of the classical r-rich EMP star CS\,31082-001 (black dots). The blue solid line represents the abundance pattern of the r-process elements in the solar system.\\
Panel B: same as Panel A, but for the three low-Sr/Ba stars studied here (open pink symbols).\\ 
In this figure, the  abundances are represented not by ratios compared to the Sun (as in the previous figures), but by the other classical notation $\rm \log \epsilon\,$: the logarithm of the number of atoms for $10^{12}$ atoms of H, scaled to the abundance of europium.
The abundance pattern of the three low Sr-Ba stars (Panel B) present many similarities with the abundance pattern of CS\,22948-027 and CS\,29497-034 (panel A).
} 
\label{pattern}
\end{figure}

\subsection{The neutron-capture elements from Sr to Pb} 

To allow a more precise comparison of the low-Sr/Ba stars to the different subsamples of metal-poor stars (EMP stars, CEMP stars), ~in Fig. \ref{scatter2} we have plotted  [Sr/Ba] vs. [Ba/Fe] for only giant stars within a narrow range of metallicity  $\rm [Fe/H]=-3.1\pm0.4$. These stars have model parameters very close to those of the low-Sr/Ba stars, and as a consequence the comparison is even more suitable.

As can be seen, the three low-Sr/Ba stars occupy a very peculiar position in this diagram, and this position is not the consequence of a difference in metallicity since the different subsamples of stars in this figure have the same mean metallicity. The low-Sr/Ba stars are Ba-rich with $\rm[Ba/Fe]>0.0$ like the EMP r-rich stars, but their Sr abundance is low with $\rm[Sr/Fe] < -0.5$, and as a consequence, the ratio [Sr/Ba] is less than the lower limit for an r-process production: [Sr/Ba]=--0.5  \citep{MashonkinaG01}.  The ratio [Sr/Ba] in the three low-Sr/Ba giants is similar to the one observed in the CEMP-s stars (Fig. \ref{scatter2}) but with a lower [Ba/Fe] ratio. 

In panel A of Fig \ref{surfe}, we compare the abundance ratios [X/Fe] (where X is any of the neutron-capture elements) in the low-Sr/Ba stars (Table \ref{abund}) to the ratios observed in r-rich stars, which are a good proxy for the typical r-process \citep{SiqueiraMelloSB13,PlezHC04}. The behaviors are  completely different, even the behavior of the Sr, Y, Zr group is different, but the main difference is in the [Pb/Eu] ratio. This is a clear indication that in our low-Sr/Ba stars, the neutron-capture elements were not formed by the r-process.\\
In panel B, we compare the abundance ratios [X/Fe] in the three low-Sr/Ba stars to the s-process production during thermal pulses in a moderately metal-poor AGB (Z=0.001) following \citet{GorielyMow00}. In this figure we shifted those computed [X/Fe] ratios by --1.4 dex for an easier comparison to the observations. In spite of the higher metallicity of the AGB model in the computations of \citet{GorielyMow00}, the abundance pattern is quite similar.\\

The similarity between the s-process computations during thermal pulses of AGB stars \citep{GorielyMow00} 
is striking (in particular the ratio [Pb/Eu]) and strongly support that the neutron-capture elements in the low-Sr/Ba stars have mainly been  formed by the s-process.

\section{Discussion}
    
\subsection{Were the low-Sr/Ba stars enriched by the ejecta of a defunct AGB star?}

In panel A of Fig. \ref{pattern}, we compare the abundance pattern of two \mbox{CEMP-s} stars CS\,22948-027 or CS\,29497-034  \citep[following][]{BarbuySS05} to the abundance pattern of a classical r-rich EMP star CS\,31082-001 \citep[][]{SiqueiraMelloSB13}. 

Compared to the abundance pattern of the r-elements in the Sun the CEMP-s stars present a strong overabundance of Ba, La, Ce, Nd, and Pb relative to Eu.\\
In panel B, we compare the abundance pattern of the three low-Sr/Ba stars to CS\,31082-001 and the solar r-process abundance pattern.
The low-Sr/Ba stars and the CEMP-s stars  have similar characteristics: strong enhancement of Ba, La, Ce, and Pb relative to Eu when compared to r-process rich stars and/or to the the solar r-process although the overabundances of the Ba-group and of Pb (relative to Eu) is more pronounced in the low-Sr/Ba stars.

The CEMP-s stars such as CS\,22948-027 or CS\,29497-034 are generally found to be binaries, and thus the distribution of their neutron-capture elements is explained by a mass transfer from an earlier AGB companion (now, a white dwarf), that has produced neutron-capture elements by the s-process. Would it be possible that the three low Sr/Ba stars in Table \ref{abund} be RGB binaries that have been enriched by an AGB companion?\\

\subsubsection{Radial velocities of the three low-Sr/Ba stars} \label{rvgi}
The radial velocities of the three low-Sr/Ba stars have been carefully measured:\\
\noindent--CS~30322-023\\ 
The radial velocity of this star has been studied in detail by 
\citet{MasseronEF06} from 12 spectra obtained between 2001 and 2006, and they conclude that there was no firm support for the binary nature of this star.\\
--CS~29493-090\\ 
On spectra obtained in 2003, \citet{BarklemCB05} measured RV=269.8 \kms~ in very good agreement with the value obtained from our 2005 spectra (Table \ref{VVr}).\\
--HE~0305-4520\\ 
On spectra obtained in 2002,  \citet{BarklemCB05} measured RV=135.3 \kms, in agreement with the value obtained from our 2006 spectrum  (Table \ref{VVr}).
%(no RV variation for CS~22950-173)

As a consequence, none of these low-Sr/Ba stars presents a clear variation in the radial velocity that could support the hypothesis that their atmosphere has accreted matter from an AGB companion.
As noted by \citet{MasseronEF06}, however,  ``the absence of evidence is not evidence of absence''. The stars could have a rather long period (although this is not favorable to a significant amount of pollution) or/and a small radial velocity amplitude. They could also have an orbit in a plane (nearly) perpendicular to the line of sight, and in this case their binarity could perhaps be detected from a variation in the astrometric position of the stars by the spatial mission Gaia \citep{deBruijne12}.

\subsection{Is it possible that the low Sr/Ba stars are AGB stars ?  }

If we exclude an external enrichment of the atmospheres of our low-Sr/Ba stars, we have to explore the possibility of an internal production of $\rm^{12}C$, $\rm^{13}C$, N, and the s-process elements with a transfer to the surface.
In low-mass stars this is sometimes possible if the star is an AGB star.

In the sample of \citet{AokiSB13}, among 260 stars  only five stars present  a low Sr/Ba ratio.  
If a star stays one tenth of its life on the RGB \citep[e.g.][]{IbenI91,PietrinferniCS04}, then our stars should stay around 1.3 Gyrs on the RGB. Following \citet{Bloeker95} the lifetime of a star of about $1M_{\odot}$, in the AGB phase, is only about $4.5\times10^{7}$ years. Therefore, we could expect to find about eight AGB stars in the \citet{AokiSB13} sample, a number compatible with the small number of low-Sr/Ba stars in the sample. 

\subsubsection{Do we see an internal pollution of the atmosphere occurring in TP-AGB stars?}

In low-mass stars an internal production of $\rm^{12}C$, $\rm^{13}C$, N and the s-process elements with a transfer to the surface occurs in thermally pulsating AGB (TP-AGB) star.
During a thermal pulse \citep[e.g.][]{KappelerGB11}, partial He burning occurs in the intershell, and after a limited number of thermal pulses, the convective envelope penetrates the top region of the He intershell and mixes newly synthesized material to the surface during the third dredge-up  \citep[see also:][]{CampbellLatt08,CampbellLK10}.

\citet{MasseronEF06} suggests that  CS~30322-023 could be a  TP-AGB star, and this would explain the observed abundances by an internal (recent)  enrichment in carbon, and in heavy elements bearing the signature of the s-process. This hypothesis is based on their having found that the gravity of CS~30322-023 is very low (\logg = --0.3), and is reinforced by the fact that $\rm H_{\alpha}$ presents a  variable emission, probably due to a stellar wind. 
However, \citet{AokiBC07} also studied this star and found a gravity  \logg = 1. They  use a {\it photometric} temperature, while \citet{MasseronEF06} determine the temperature by requiring that the abundance deduced from the \Feu ~lines  be independent of the excitation potential of the line, including in their computations   \Feu ~lines with very low excitation potentials, so   sensitive to NLTE effects \citep{CayrelDS04}. This difference in temperature has no significant effect on the abundance ratios, but does induce a difference in gravity (since gravity is determined in both cases from the ionization balance between \Feu~ and \Fed).
It will be very interesting to obtain a direct measurement of the luminosity (and thus of \logg) of this star from a future precise parallax provided by the spatial mission Gaia, defining the precise evolution status of CS~30322-023.

CS\,29493-090 and  HE\,0305-4520 have similar abundance ratios as CS~30322-023, but they are hotter with a higher gravity (\logg=1.3, see Table \ref{models}) and do not show emission in $\rm H_{\alpha}$.
From Fig. 8 of \citet{MasseronEF06}, they cannot be TP-AGB, but they could be early-AGB. Could it be possible that, already at this phase of the stellar evolution, neutron-capture elements be produced in the stellar interior and brought to the surface by a deep mixing?

\subsection{Do we see the s-process production of the core He flash in the low-Sr/Ba stars ? }
In metal-rich stars, during the helium flash at the tip of the RGB, convection cannot break the H-burning shell and bring the products of the helium-burning nucleosynthesis  to the surface, since the active H-burning shell provides an entropy barrier against mixing. However \citet{CampbellLK10} have shown that in hyper metal poor stars ($\rm[Fe/H] < 5.0$), this barrier can be broken (helium flash induced mixing), and the material, processed in the helium burning shell, is carried to the stellar surface by a dredge-up event.
Our low-Sr/Ba stars are not as metal-poor as the model computed by \citet{CampbellLK10}, but it would be interesting to check if in stars with $\rm[Fe/H] = -3.0$  this process could sometimes not be, in some particular conditions, efficient.
%In the model computed by  \citet{CampbellLK10}, $\rm[C/Fe]\simeq [Ba/Fe]$ similar to the observations in low Sr/Ba stars. On the contrary the Campbell et al. model predicts the same over abundance of N ($\rm[N/Fe]\simeq [Ba/Fe]$) at variance with what is observed in low-Sr/Ba stars ($\rm[N/Fe]\simeq [Ba/Fe]+1.5$)}. 

\section {Conclusion}   

We analyzed four extremely metal-poor stars (EMP)  for which anomalously low-[Sr/Ba] ratios were reported in the literature (lower than the ratio of the typical main r-process).  One of them (the only turnoff star) turns out to have a "normal"  Sr/Ba ratio (and was discarded from this work). The three remaining stars are all cool evolved giants. The abundance ratios of the elements lighter than the iron peak are generally not compatible with the classical EMP stars (Figs. \ref{cpn} to \ref{mg}). Compared to the classical EMP stars these low-[Sr/Ba] stars 
present an enhancement of C+N, Na, and Mg.
From their temperature and gravity they can be evolved RGB stars or AGB stars.

The moderately high C/Fe ratio measured in the three giants does not make them CEMP stars following the definition of \citet{BeersChris05}, but we showed that these cool giants ($\rm T_{eff} \leq 4800K$) with a low gravity ($\log g <1.5$) present the characteristics of stars having undergone a deep mixing (in the RGB phase). In this case, part of their initial C would have been transformed into N during their evolution, therefore the carbon abundance in the cloud that formed the star was probably much higher than the value measured today in their atmosphere;  also the stars could be CEMP stars (following the definition of \citealp{BeersChris05}) at their birth. For evolved giants, it would be more suitable to define the CEMP type from the C+N abundance of the star rather than from their C abundance alone: for example, CEMP stars could be defined (Fig. \ref{cpn}) by $\rm[(C+N)/Fe] > 0.8$. 

In the CEMP-s stars, the low Sr/Ba ratio is generally associated to $\rm[Ba/Fe] >1$  (Fig. \ref{scatter2}), and to high abundances of Ba and Pb relative to Eu. The radial velocity  of these stars is variable \citep{StarkenburgSC14},  and their abundances are interpreted as a superficial pollution (by a mass transfer from a now extinct AGB companion) providing C, N, and the neutron-capture elements produced by the s-process.

The abundance pattern of the neutron-capture elements in our three low-Sr/Ba stars is slightly different from the pattern of CEMP-s stars:  the abundances of the neutron-capture elements relative to europium is larger (Fig. \ref{pattern}). The distribution of the ratios [X/Fe] (where X is a neutron-capture element) vs. the atomic number Z is very similar to the predictions of \citet{GorielyMow00} or \citet{CampbellLK10} for a pure s-process.
\\
What is the origin of the enrichment by the s-process?\\
\\
We consider different possibilities:\\
\\
$\bullet$ The stars could be classical RGB stars that would have undergone an external enrichment by a now defunct AGB star, as stated for the CEMP-s stars. In this case, we should expect, at least statistically, a variation in the radial velocity of these stars. No variation is present clearly in any of our three low-Sr/Ba stars but a further monitoring of their radial velocity would be very useful for a firm decision.\\

\noindent $\bullet$ In the absence of  radial velocity variations, it is interesting to investigate whether autopollution is a valuable alternative solution.

A He shell flash may be considered, but it would occur when the stars are already in the TP-AGB phase, and it cannot occur in an early-AGB. As a consequence, this process does not seem able to explain the peculiar abundances of the low-Sr/Ba stars, which have a relatively high gravity ($\log g \sim 1.3$).

The stars may have undergone a dual core flash at the tip of the RGB, but this mechanism is predicted only in hyper low metallicity stars \citep{CampbellLatt08,CampbellLK10} and a higher C/N ratio would be expected. Moreover, the surface composition is expected to remain unchanged from the core flash dredge-up until the beginning of the TP-AGB phase.\\

\noindent More definitive conclusions would require both 
(i) a check of a possible binarity of the three low-Sr/Ba stars from new precise measurements of the radial velocities, and also of the astrometric position of the stars (space astrometry mission Gaia). 
(ii) measurements of the luminosity (distance) of the stars with Gaia, to derive their evolution status more precisely: early-AGB or TP-AGB. Maybe in the future, asteroseismology could help to distinguish between RGB and AGB stars.

\begin {acknowledgements} 
We  made use of SIMBAD operated by the CDS in Strasbourg, of the ADS Service
(SAO/NASA), and of the arXiv Server (Cornell University). The authors acknowledge the support of Observatoire de Paris-GEPI and  the support of the CNRS (PNCG and PNPS). E.C. is grateful to the FONDATION MERAC for funding her fellowship. 
\end {acknowledgements}

\bibliographystyle{aa}
{}
\end{document}